\documentclass[useAMS,usenatbib,usegraphicx]{mn2e}
\bibpunct{(}{)}{;}{a}{}{,}

\input epsf.sty

\title[Iron $K_\alpha$ fluorescence line and Compton shoulder]{ 
Models of the iron $K_\alpha$ fluorescence line and the Compton shoulder in 
irradiated accretion disk spectra}

\author[A. R{\'o}\.za{\'n}ska, J. Madej ]{
A. R{\'o}\.za{\'n}ska$^{1}$\thanks{E-mail: agata@camk.edu.pl (AR); 
jm@astrouw.edu.pl (JM)}, J. Madej$^{2}$  \footnotemark[1] \\
$^{1}$ Copernicus Astronomical Center, Bartycka 18, 00-716 Warsaw, Poland \\
$^{2}$ Warsaw University Observatory, Al. Ujazdowskie 4,
       00-478 Warsaw, Poland }

\begin{document}
\date{Accepted .................................. }
\pagerange{\pageref{firstpage}--\pageref{lastpage}} \pubyear{2007}

\maketitle
\label{firstpage}

\def\te{T_{\rm eff}}

\begin{abstract}
We present a full set of model atmosphere equations for the 
accretion disk around a supermassive black hole irradiated by hard X-ray
lamp of power-law spectral distribution. Model equations
allow for multiple Compton scattering of radiation on free electrons,
and for large relative photon-electron energy exchange at the time of
scattering. We present spectra in specific intensities
integrated over the disk surface.
Theoretical outgoing intensity spectra show soft X-ray excess below
1 keV, and distinct $K_\alpha$ and $K_\beta$ fluorescent lines of iron.
We demonstrate the existence of the Compton Shoulder and claim that
it can contribute to the asymmetry and equivalent widths of some observed 
Fe $K_\alpha$ lines in AGN.
Our models exhibit the effect of limb-brightening in reflected X-rays. 
\end{abstract}


\section{Introduction}

Observations of many active galactic nuclei (AGN) directly show
the existence of an X-ray source in the innermost region of AGN,
which emits radiation with power-law spectrum extending to above 100
keV \citep{guainazzi1999, done2003}.
The exact location of the source is still unclear, nevertheless, such hard
radiation strongly affects other emitting regions including nearby accretion
disk. Effects of the redistribution of hard X-rays into the soft X-ray and
ultraviolet domains are frequently observed in many  AGN starting with the
paper \citet{pounds1990} up to the recent paper \citet{nandra2007}. 

The most important spectral feature indicating that the reflection
occurs from an accretion disk, is a fluorescent iron $K_\alpha$ line.
Close to the supermassive black hole, the emission iron line  profile
is relativistically broadened and skewed \citep{fabian1989,
reynolds1997}, which together with a circular gas movement in 
a disk, makes final profile asymmetric with a strong red wing. 

The detection of iron line profile was a goal of several X-ray satellites,
and the results depend on the resolution of a particular instrument. 
The best known object  with broad iron $K_\alpha$ line, MCG-6-30-15 was observed by 
{\it ASCA} \citep{tanaka1995, iwasawa1999}. In the {\it Chandra} 
and {\it XMM} era, line became narrow or contained both broad and narrow
component \citep{reeves2004, petrucci2007}. Recently, more often authors are
reporting a detection of ``iron line complex'' additionally affected by 
intrinsic warm absorber \citep{yaqoob2004, risaliti2005, matt2006}.

The most recent data coming from {\it Suzaku} look very promising 
\citep{miniutti2007, reeves2007, yaqoob2007}. The iron line complex 
contains at least two components: narrow Gaussian at 6.4 keV, and broad red
wing, the last one usually fitted with a disk like shape, but also with 
Compton shoulder in the work of \cite{reeves2007}.  Additionally iron 
$K_\beta$ line at energy 7.057 keV is clearly seen in {\it Suzaku} AGN 
\citep{reeves2007, markowitz2007, yaqoob2007}.

In this paper we present quite consistent and numerically exact modeling
of the iron line complex originating from an accretion disk atmosphere 
irradiated by hard X-rays with power-law spectral distribution. 
Our computations of the vertical structure and outgoing spectrum of the 
disk are based on rigorous method of the theory of stellar atmospheres 
\citep{madej1991, madej2000a, madej2000b}. All existing opacity sources 
both true absorption and Compton scattering are treated equivalently in 
the single code fully described in 
\citet[hereafter MR04]{madej2004}.
Adopting atmospheric computations for an accretion disk around supermassive
black hole, we solve the structure and obtain outgoing spectra for eight
neighboring rings.  The final spectrum presented in specific intensity 
scale for different aspect angles is integrated over radii and
presented at the source frame.  No kinematic special relativity
effects  are included in our model. 

Our radiative transfer equation includes effects of multiple Compton 
scattering of radiation on free electrons in relativistic thermal motion,
and rich set of bound-free and free-free opacities. We allow for a large
relative photon-electron energy exchange at the time of Compton scattering
and, therefore, are able to reconstruct Compton scattering of photons with
energy approaching or even exceeding the electron rest mass. 
In this paper the opacities are supplemented by formulae describing emission
of fluorescent $K_\alpha$ and $ K_\beta$ lines of low ionized iron.

The structure of the paper is as follows: Sec.~\ref{sec:mod}
describes sample accretion disk model, and Sec.~\ref{sec:atm}
presents most important equations used in our code. 
Results are presented in Sec.~\ref{res:temp} and Sec.~\ref{res:line} 
and summarized in Sec.~\ref{sec:discussion} and Sec.~\ref{sec:sum}.

\section{Model of a sample accretion disk}
\label{sec:mod}

We consider an exemplary  model of the  accretion disk around supermassive
black hole located at the center of AGN. Mass of the black hole equals 
$M_{BH}=10^7 M_\odot$. We divide our disk onto 8 concentric rings situated
at different distances from the center. We assume constant moderate value
of accretion rate in each ring, equal $\dot m=0.03$ in units of Eddington
accretion rate with accreting efficiency $\eta=1/12$ suitable for the
Schwarzschild black hole. 

At the first step, at each radius, we computed vertical structure of 
non-irradiated disk using the method described in \citet{rozanska1999}.
We integrated equations of the disk vertical structure (equation of the 
hydrostatic equilibrium, equation of state, equation of local energy 
generation, transfer in diffusion approximation) assuming that viscosity
is proportional to the total pressure, i.e. $P_{tot}=P_{rad}+P_{gas}$.
As the result, for each ring, we derived effective temperature and vertical
gravity which affects atmosphere, and we adopted them for actual computations. 
All disk parameters used in further computations are summarized in 
Tab.~\ref{tab:disk} column 3 and 4. 

At the next step, we assume that the disk is irradiated by an external point
like X-ray lamp, located above the first innermost ring, 
$r_1=3.48 \, r_{Schw} $, at the height $h_{l}=5\, r_{Schw}$.
Irradiation spectrum is in the form of power-law with exponential low and
high energy cut-off (see Sec.~\ref{sec:exirr}).

The code by \citet{rozanska1999} predicted the value of gravity, 
$\log g_m$, for each ring of the non-irradiated disk. When external 
irradiation was applied, the atmosphere always lost its hydrostatic 
equilibrium. We had to increase the value of surface gravity by a factor 
of two or more to restore hydrostatic equilibrium  in those rings, 
where equilibrium probably may not exist. There is no firm physical 
justification for such a step, of course.
Values of gravity adopted for the advanced radiative transfer computations,
$\log g_{irr}$, are listed in the fifth column of Tab.~\ref{tab:disk}. 

We take into account three different chemical compositions of the 
accretion disk presented in  Table~\ref{tab:elem}. The first composition, 
Comp. I, consists only of hydrogen, helium and iron in proportions close
to solar values. Comp. II is just the solar chemical composition for only ten 
most abundant elements given by  \citet{chaisson96}. We also considered the
model of solar composition with doubled number abundance of iron named as
Comp. III.

In this paper, we have calculated our canonical model of the disk with
chemical composition Comp. I, and the power-law irradiation with spectral
index $\alpha_{X}=0.9$, and X-ray luminosity $L_X=10^{43}$ erg s$^{-1}$.  
For chemical composition, Comp. I, we compared this canonical model
to the case of power-law with the same spectral index, but different 
luminosities: $L_X=10^{42}$ erg s$^{-1}$ and $L_X=10^{44}$ erg s$^{-1}$,
and to the case of power-law with the same luminosity but different spectral
indices: $\alpha_{X}=0.6$ and $\alpha_{X}=1.2$. 
We also compared the structure and  spectrum of our canonical model 
to models with the luminosity $L_X=10^{43}$ erg s$^{-1}$,
and spectral index $\alpha_{X}=0.9$, but with other chemical compositions:
Comp. II and Comp. III. 
In all computed models the
power-law radiation extends between $h \nu_{min} = 0.1 $ and $h \nu_{max} = 100$
keV. 

\begin{table}
\caption{ Parameters of the disk for $M_{BH}=10^7 M_\odot$,
$\dot m=0.03$. }
\begin{tabular}{|c|c|c|c|l|}
\hline \hline
  ring No. &$r/r_{Schw}$ &$T_{\rm eff}$ &$\log g_m$ &$\log g_{irr}$ \\ 
\hline
   1  &  3.478E+00 &  1.018E+05 & 5.230E+00  & 5.5300  \\
   2  &  4.134E+00 &  1.073E+05 & 5.274E+00  & 5.5740  \\
   3  &  5.204E+00 &  1.020E+05 & 5.170E+00  & 5.4700  \\
   4  &  6.552E+00 &  9.238E+04 & 5.001E+00  & 5.3010  \\
   5  &  8.248E+00 &  8.181E+04 & 4.801E+00  & 5.1010  \\
   6  &  1.038E+01 &  7.152E+04 & 4.582E+00  & 5.1820  \\
   7  &  1.385E+01 &  5.976E+04 & 4.294E+00  & 5.594   \\
   8  &  1.847E+01 &  4.950E+04 & 3.996E+00  & 5.596   \\
\hline \hline
\end{tabular}
\label{tab:disk}
\end{table}

\begin{table}
\caption{Three chemical compositions used in our calculations. 
Entries of the table display number abundances of elements 
relative to hydrogen. Three rows at the bottom present mass 
abundances X, Y, and Z for all three composition.}
\begin{tabular}{|c|c|c|c|}
\hline \hline
      &Comp. I & Comp. II & Comp. III  \\ 
\hline
  H  & 1.0 & 1.0 & 1.0  \\
 He  & $1.1 \times 10^{-1}$  & $9.54 \times 10^{-2}$  & $ 9.54 \times 10^{-2} $ \\
 C & - &  $4.72 \times 10^{-4}$ & $4.72 \times 10^{-4}$  \\
 N  & - & $9.65 \times 10^{-5}$ & $9.65 \times 10^{-5}$ \\
 O  & - & $8.55 \times 10^{-4}$ & $8.55 \times 10^{-4}$ \\
 Ne & - & $3.84 \times 10^{-5}$ & $3.84 \times 10^{-5}$ \\
 Mg  & - & $4.17 \times 10^{-5}$ & $4.17 \times 10^{-5}$ \\
 Si & - & $4.94 \times 10^{-5}$ & $4.94 \times 10^{-5}$  \\
 S & - & $1.64 \times 10^{-5}$ & $1.64 \times 10^{-5}$   \\
 Fe & $3.7 \times 10^{-5}$ & $3.29 \times 10^{-5}$ & $6.58 \times 10^{-5}$ \\
 \hline
 X & 0.695 & 0.712 & 0.711  \\
 Y & 0.304 & 0.270 & 0.269  \\
 Z & $2.066 \times 10^{-2}$ & $2.624 \times 10^{-2}$ &$2.808 \times 10^{-2}$ \\
\hline \hline
\end{tabular}
\label{tab:elem}
\end{table}

\section{Equations of an irradiated atmosphere}
\label{sec:atm}

The set of equations presented and solved here represents a direct extension
of the equations presented in MR04. Now we added
terms which correspond to the fluorescent $K_\alpha$ and $K_\beta$ lines
of iron, adjusted to the accretion disk geometry and assumed that the external
irradiation from a point X-ray lamp has the power-law spectral distribution.

\subsection{The equation of transfer}

The equation of transfer for the specific intensity  $I_\nu$ at frequency
$\nu$ was written in plane-parallel geometry on the geometrical depth 
scale $z$
\begin{equation}
\mu \, {dI_\nu \over{\rho \, dz}}=j_\nu-(\kappa_\nu+\sigma_\nu) I_\nu \, ,
\label{eq:int}
\end{equation}
where $j_\nu$, $\kappa_\nu$ and $\sigma_\nu$ denote frequency dependent 
emission, absorption and scattering coefficients for 1 gram, respectively.
In this paper we use the LTE (local thermodynamic
equilibrium) absorption $\kappa_\nu$, whereas coefficients of emission
$j_\nu$ and scattering 
$\sigma_\nu$ include nonLTE terms.

Eq.~\ref{eq:int} can be written on the monochromatic optical depth, with 
$d\tau_\nu = - (\kappa_\nu + \sigma_\nu) \rho \, dz$, then
\begin{equation}
\mu \, {dI_\nu \over {d\tau_\nu}} = I_\nu  - {j_\nu \over {\kappa_\nu +
  \sigma_\nu}} = I_\nu - S_\nu \, ,
\end{equation}
where $S_\nu$ is the frequency dependent source function.

Emission coefficient $j_\nu$ is the sum of three terms, 
$j_\nu = j_\nu^{th} + j_\nu^{sc} + j_\nu^{fl}$, which represent thermal
emission, Compton scattering emission and the emission in iron fluorescent
lines, respectively.

\subsubsection{Thermal emission}

Coefficient of thermal emission $j_\nu$ in LTE is proportional to the
Planck function $B_\nu$
\begin{equation}
j_\nu^{th} = \kappa_\nu \, B_\nu \, .
\end{equation}
The coefficient of true absorption $\kappa_\nu$, is the sum of
bound-free absorption from numerous levels of atoms 
and ions for all elements, plus free-free absorption from all ions.  

We also included absorption of 4 lowest lines of fundamental series 
of helium-like iron and of similar 4 lowest lines of hydrogen like iron,
all formed in LTE by assumption.

\subsubsection{External irradiation}
\label{sec:exirr}

We defined two angle-averaged variables, which were derived from the 
specific intensity $I_\nu^{ext}$ of the external irradiation penetrating
the atmosphere to the depth $\tau_\nu$
\begin{equation}
U_\nu (\tau_\nu) = {1\over {4\pi}} \int_\Omega I_\nu^{ext}(\omega,\tau_\nu)
  \exp \, (-\tau_\nu/\mu_\omega) \, d\omega \, ,
\label{equ:meanu}
\end{equation}
\begin{equation}
V_\nu (\tau_\nu) = {1\over {4\pi}} \int_\Omega I_\nu^{ext}(\omega,\tau_\nu)
  \, \mu_\omega \exp \, (-\tau_\nu/\mu_\omega) \, d\omega \, ,
\label{equ:meanv}
\end{equation}
where $\mu_\omega$ stands for the cosine of zenithal angle of the running
direction $\omega$. 
The external intensity from the point-like lamp is
emitted in the form of power-law with spectral index $\alpha_X$:
\begin{equation}
I_{\nu}^{ext}=A \, \nu^{- \alpha_X} \, exp\left(- \frac{\nu}{\nu_{max}}\right)
\, exp\left(- \frac{\nu_{min}}{\nu} \right),
\end{equation}
and  normalized to the luminosity of the source, $L_X$:
\begin{equation}
A  =  \frac{ L_X}{4 \pi r_l^2 \left[  \int_{\nu_{min}}^{\nu_{max}}
\nu^{- \alpha_X} \, exp\left(- \frac{\nu}{\nu_{max}}\right)  
   \, exp\left(- \frac{\nu_{min}}{\nu} \right) d\nu \right]} .
\end{equation}
The distance from an X-ray source depends on the ring radius $r$
in the casual relation $r_l^2=h_l^2+r^2$.
The luminosity, spectral index, and cut-off limits of irradiating spectrum
are free parameters of our model, all of them described in Sec.~\ref{sec:mod}.

\subsubsection{Compton scattering emission}

Expressions for the Compton scattering terms and emission coefficient in
an irradiated stellar atmosphere were derived in MR04 and are valid
in a disk atmosphere without any changes

\vspace{-2mm}
\begin{eqnarray}
j_\nu^{sc} & = & \sigma_\nu J_\nu - \sigma_\nu J_\nu \int_0^\infty 
  \Phi_1 (\nu, \nu^\prime)\, d\nu ^\prime \cr & & + \sigma_\nu \int_0^\infty
  (J_{\nu^\prime}+U_{\nu^\prime})\, \Phi_2(\nu, \nu^\prime)\, d\nu^\prime\, .
\end{eqnarray}
In the above equation variable $U_\nu$ denotes the angle-averaged intensity
of the external irradiation (Eq.~\ref{equ:meanu}). 

Compton scattering cross sections were computed following the paper by
\citet{guilbert1981}. Functions $\Phi_1$ and $\Phi_2$ are properly weighted 
angle-averaged Compton redistribution functions for photons both incoming or outgoing
of frequency $\nu$ after scattering in thermal electron gas, cf. MR04.

\subsubsection{Iron fluorescence emission}

Fluorescence of low-ionized iron gas was approximated by two emission lines,
$K_\alpha$ and $K_\beta$. Therefore,
\begin{equation}
j_\nu^{fl} = E_\alpha^{fl} \, \varphi_\nu^\alpha + 
   E_\beta^{fl} \varphi_\nu^\beta \, ,
\end{equation}
where $E_\alpha^{fl}$ and $E_\beta^{fl}$ denote the integrated intensity
of $K_\alpha$ and $K_\beta$ emission lines, respectively. Frequency
dependent variables $\varphi_\nu^\alpha$ and $\varphi_\nu^\beta$ define
profiles of fluorescent lines, both normalized to unity.

Energy emitted in $K_\alpha$ and $K_\beta$ lines was derived by absorption
of hard continuum X-rays from the radiation field  penetrating the disk
atmosphere. Hard X-ray photon interacting with neutral or low-ionized iron
most probably ionize and remove electron from the innermost $K$ shell, and
then remaining hole is filled by another electron falling from $L$ shell 
($K_\alpha$) or $M$ shell ($K_\beta$ transition). 
The integrated emissivity for Fe $K_\alpha$ fluorescent line is given by:
\begin{equation}
E_\alpha^{fl} = Y \times h\nu_0 \int_{\nu_0}^\infty
  {\alpha_\nu^{\rm iron} \over {h\nu}} (J_\nu + U_\nu) \, d\nu \, ,
\end{equation}
where we set $Y=0.34$ \citep{krause1979}. Variable $\alpha_\nu^{\rm iron}$
is the bound-free absorption coefficient for ionization from $K$ shell
of iron counted for 1 atom. Similar expression holds for the integrated
intensity $E_\beta^{fl}$.

The fluorescence yield, $Y$, defines the fraction
of the energy of hard X-rays absorbed by iron atoms which was reemitted as
photons in $K_\alpha$ line. The remaining energy of absorbed X-rays,
$1-Y$, was spent for the release of Auger electrons.

The iron $K_\alpha$ fluorescent line is a doublet line,
and such a structure was reproduced by our code.   
We set the central energies for $K_{\alpha_1}$ and
$K_{\alpha_2}$ lines to 6.404 keV and 6.391 keV, respectively. 
Natural widths (FWHM) of both lines, 2.7 eV ($K_{\alpha_1}$) and 3.3 eV 
($K_{\alpha_2}$) were taken from \citet{krause1979}.
The $K_\beta$ line was approximated by a singlet line at the energy
7.057 keV. We set the natural width of the line to 2.5 eV arbitrarily.

Integrated emissivity $E_\alpha^{fl}$ was divided in the proportion 2:1 
between $K_{\alpha_1}$ and $K_{\alpha_2}$ components of the doublet.
 Following 
\citet{basko1978} we quite arbitrarily assumed that the integrated
intensity $E_\beta^{fl} = 0.1 \, E_\alpha^{fl}$.
Opacity profiles of all three lines were set to Voigt functions with 
depth-dependent parameters describing natural and thermal broadening.

\subsection{Equation of state}

We used here the LTE equation of state, where
the number of free electrons and all ionization
populations of iron was determined by the Saha ionization equations.

In the case of an atmosphere irradiated by strong non-local radiation field
the assumption of LTE yields inaccurate ionization and excitation populations.
For instance we get inaccurate populations of iron atoms with various
configurations of electrons in the outermost M and N shells.

This is the reason that we did not discuss in our paper
intensities for the
spectral features of energies below the soft X-ray excess in the outgoing
spectra, where the nonLTE configuration of outermost bound electrons in various
atoms is of vital importance.

However, we believe that the population of iron atoms with filled innermost
K and L shells is always equal to the total number of iron atoms with 8 or
more bound electrons, and does not depend on the LTE/nonLTE details of 
electron populations in M and N shells.
As the result we believe that have obtained reasonable intensities for both
fluorescent lines $K_{\alpha}$ and $K_{\beta}$ of low-ionized iron. 
The lines were considered strictly in nonLTE.

\subsection{Equations of hydrostatic and radiative equilibrium}

\subsubsection{Hydrostatic equilibrium}

We solve the static equation of equilibrium for gas pressure:
\begin{equation}
  { dP_{gas} \over {d\tau}} = {g \over {(\kappa + \sigma )_{std} }} -
  { dP_{rad} \over {d\tau}} \, ,
\end{equation}
which takes into account gradient of radiation pressure $P_r$ at the level
$\tau$. Expanded form of the equation is
\begin{equation}
  { dP_{gas} \over {d\tau}} = {g \over {(\kappa + \sigma )_{std} }} -
  { 4\pi \over c} \int_0^\infty \eta_\nu (H_\nu - V_\nu) \, d\nu \, ,
\label{equ:hydr}
\end{equation}
where $\eta_\nu = (\kappa_\nu+\sigma_\nu)/(\kappa+\sigma)_{std} $.
Eq. \ref{equ:hydr} neglects possible vertical motion of matter which can
be caused by the differential rotation of the disk. 

In addition we assumed that the vertical
gravitational acceleration in the atmosphere of a given ring is constant
i.e. its dependence on the height above the disk equatorial plane was
neglected. The value of acceleration was computed from the
solution of the disk model presented in Sec.~\ref{sec:mod}. 

\subsubsection{Radiative equilibrium}

Constraint of radiative equilibrium requires that 
\begin{equation}
\int_0^\infty H_\nu (\tau_\nu) \, d\nu - \int_0^\infty V_\nu (\tau_\nu)
   \, d\nu = {\sigma_R T_{\rm eff}^4 \over {4\pi }}
\label{equ:rad1}
\end{equation}
at each depth level of $\tau_\nu$ in the atmosphere. Radiative constant
$\sigma_R = 5.66961 \times 10^{-5}$ (cgs units). In this paper, we assumed
that whole flux from an accretion disk was generated below an atmosphere,
and we do not solve local energy generation via viscosity. 

The above equation subsequently was used to determine distribution of 
temperature in the disk atmosphere at a given ring. Transformation of
Eq. \ref{equ:rad1} to the expression for temperature corrections 
$\delta T$ was done by the linearization technique.
The equation of transfer including temperature corrections was solved
using the technique of variable Eddington factors \citep{mihalas1978}.
Details of the full set of equations used in our computations and the 
method of solution by partial linearization are presented in MR04.

\section{Computational results}
\label{res:temp}

Our computer code allowed us to compute the structure of disk atmospheres
over very large range of electron scattering optical depth starting from
$\tau_{es}= 10^{-8}$ up to $\tau_{es}= 10^{5}$. Furthermore, we were able
to reproduce the overall continuum spectrum from deep infrared of 0.4 eV
up to hard X-rays of 400 keV. We present our spectra as an energy dependent 
outgoing specific intensities, $I_\nu$, which are suitable for disk geometry. 
We reject presentation of monochromatic fluxes since they are relevant only
to geometry of a spherical star.

\subsection{Temperature structure}

\begin{figure}
\epsfxsize=8.8cm \epsfbox[100 240 400 690]{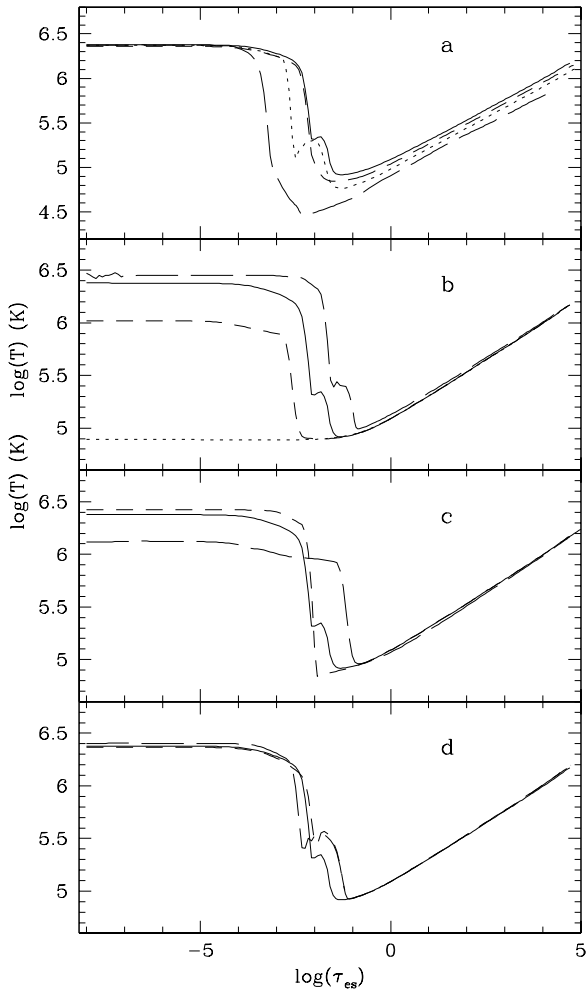}
\caption{Comparison of temperature structure for different models.
In panel (a) our canonical model is presented for 
ring 1 - solid line, ring 4 - short-dashed line, ring 5 - dotted line, and
ring 8 - long-dashed line.
In panel (b) canonical model (solid line), is compared with 
models of different luminosities:  $L_X=10^{44}$
erg~s$^{-1}$ - long-dashed line, $L_X=10^{42}$ erg~s$^{-1}$ -
short dashed line,  and dotted line - model with no irradiation. 
The same canonical model (solid line) in panel (c) is compared with the model
of $\alpha_X=0.6$ - short dashed line, and $\alpha_X=1.2$ - long dashed line.
Panel (d) shows the canonical model (solid line), 
compared with the structures of the atmosphere illuminated by the same 
radiation: $L_X=10^{43}$ erg~s$^{-1}$, $\alpha_X=0.9 $,
but for Comp.II - short dashed line, and Comp. III - long dashed
line. Only first rings are presented for all models in panels (b), (c) and (d).}
\label{fig1:struc}
\end{figure}

\begin{figure}
\epsfxsize=8.8cm \epsfbox[100 340 400 690]{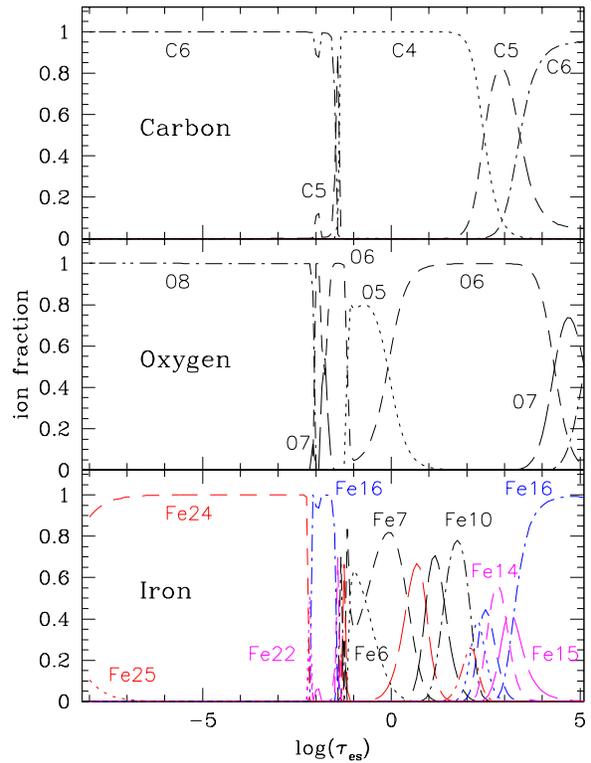}
\caption{Ionization structure of three main elements for the model with 
solar chemical composition (Comp.II see Tab.~\ref{tab:elem}).
Panel (a) shows ionization fraction of carbon ions, panel (b) - oxygen ions, 
and panel (c) - iron ions. 
 }
\label{fig:ion}
\end{figure}

Fig.~\ref{fig1:struc} represents the temperature structure of various
irradiated models. The canonical model for several rings is presented in 
panel (a). In panel (b) our canonical model is compared with models
of the same spectral index, but with different X-ray luminosities, 
while in panel (c) with models of the same X-ray luminosity but
with different spectral indices.  Panel (d) shows the structure 
of the disk irradiated by the power-law with $\alpha_X=0.9$
and luminosity  $L_X=10^{43}$ erg~s$^{-1}$ but for various 
chemical compositions. In panels (b), (c), (d) we  
show temperature structure only for the innermost ring number 1. 
 
In all computed cases we note the existence of very hot outermost 
atmospheric layer which is caused by external irradiation 
\citep{nayakshin2000, ballantyne2001, hubeny2001, rozanska2002}.
The effective temperature of all rings resulted from the accretion rate 
and non-zero viscosity, is slightly less or equal to $10^5$ K. However, 
the temperature of the outermost hot skin approaches $3 \times 10^6$~K. 
For deeper layers external irradiation practically does not change 
the temperature of the gas. 

The surface temperature is determined by the balance between Compton
heating caused by hard X-ray photons from the external illumination and 
inverse Compton cooling by the low-energy reprocessed radiation and thermal
radiation from the disk below. Hence the actual surface temperature is lower
than the Compton temperature due to the illumination alone.

In our canonical model, the temperature of the hot skin is almost the same
for all rings (panel (a))  and apparently it does not change with the
distance from the central black hole for a given power-law hardness and
the luminosity of the external irradiation.
We demonstrate, that the surface temperature  depends on the
X-ray luminosity (panel (b)) and the hardness of spectrum (panel (c)).
For softer power-law, $\alpha_X=1.2$, Compton heated skin has lower
temperature, but transition from hot to colder layer occurs deeper in the
atmosphere (long dashed line in panel (c)).

The transition between hot and cold zone is very steep,
geometrically thin, and depends on the initial conditions of
a particular ring. For the first ring the transition occurs 
deeper in an atmosphere than for the outermost eight ring (panel (a)).  
The shape of the transition is caused by ionization on heavy elements
and sometimes possesses small horizontal branch, which is
consistent with results from previous photoionization computations
\citep{nayakshin2001}.
The location of such horizontal branch changes
for different chemical compositions as shown in panel~(d).  

The temperature profile of an irradiated disk
causes the ionization structure of the atmosphere to exhibit
two distinctly separated regions. The surface high temperature zone
is just the region where heavy elements are in highly ionized states.
Below that all elements change into low ionization states.
We have illustrated this behaviour in Fig.~\ref{fig:ion}, prepared
for the ring 1, irradiated by power-law with $\alpha_X=0.9$
and luminosity  $L_X=10^{43}$ erg~s$^{-1}$, 
with solar chemical composition, Comp. II, (see: Tab.~\ref{tab:elem}).
The ion number denotes the number of missing electrons in particular
atom.   In the outermost hot layer, oxygen and carbon are completely ionized, 
while the most abundant iron, Fe24, is in the form of helium-like.

\subsection{Outgoing intensity spectra}

\begin{table}
\begin{center}
\caption{Description of lines in  Figs.~\ref{fig2:spec}-~\ref{fig6:fe}}
\label{tab:cosin}
\begin{tabular}{cccccc} 
\hline \hline 
type of a line & $\cos(i)$ & $i$ \\\hline
solid black  & 0.9801 &  $11.4^{\circ} $ \\
short-long dashed &  0.8983 &  $26.1^{\circ} $ \\
long dashed dotted   &  0.7628 &  $40.3^{\circ} $\\
short dashed dotted   &  0.5917 &  $53.7^{\circ} $\\
dotted   &  0.4083 & $ 65.9^{\circ} $ \\
short dashed   &  0.2372 &  $76.3^{\circ} $ \\
long dashed  &  0.1017 &  $84.2^{\circ} $\\
solid red &  0.0199 & $88.9^{\circ} $\\
\hline \hline
\end{tabular}
\end{center}
\end{table}

\begin{figure}
\epsfxsize=8.8cm \epsfbox[100 240 400 690]{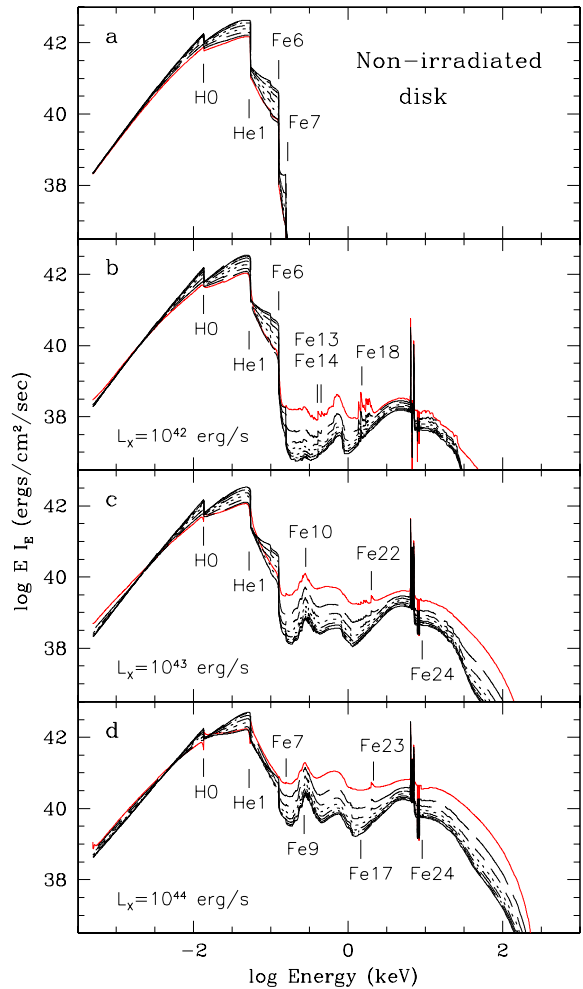}
\caption{Outgoing spectra from the disk integrated over radii for eight 
different aspect angles (see Tab.~\ref{tab:cosin}). Red solid line represents
edge on disk, while black solid line represents disk seen face on.
Panel (a) shows spectra of the disk with no irradiation.  
Panels (b), (c), (d) show models with the same $\alpha_X=0.9$, but for three
different luminosities $10^{42}$, $10^{43}$ and $10^{44}$ erg~s$^{-1}$, 
respectively.  The most important ions producing the key spectral 
features have been indicated.}
\label{fig2:spec}
\end{figure}

\begin{figure}
\epsfxsize=8.8cm \epsfbox[100 340 400 690]{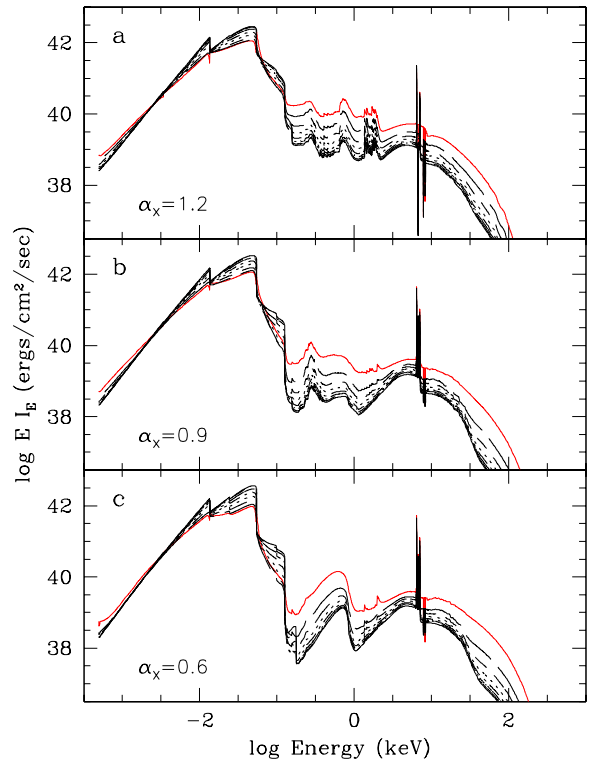}
\caption{Outgoing spectra from the disk integrated over radii for eight 
different aspect angles (see Tab.~\ref{tab:cosin}). Red solid line represents
edge on disk, while black solid line represents disk seen face on.  
Panels (a), (b), (c) show models with the same $L_X=10^{43}$ erg~s$^{-1}$, but
for three different spectral indices $1.2$, $0.9$ and $0.6$, respectively. }
\label{fig3:bb}
\end{figure}

\begin{figure}
\epsfxsize=8.8cm \epsfbox[100 340 400 690]{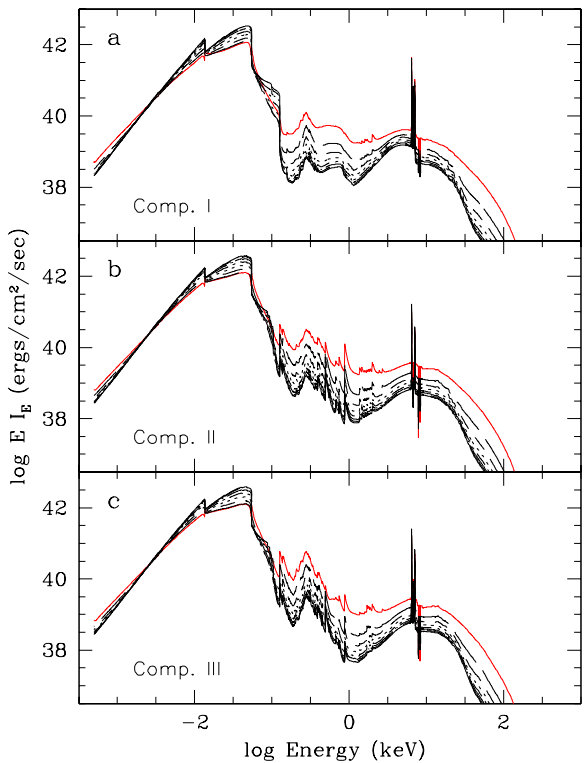}
\caption{Outgoing spectra from the disk integrated over
radii for eight different aspect angles (see Tab.~\ref{tab:cosin}). 
Red solid line represents edge on disk, while black solid line represents
disk seen face on. Panels (a), (b), (c) show models
with the same $\alpha_X=0.9$, and  $L_X=10^{43}$ erg~s$^{-1}$, 
but for three different chemical compositions: Comp. I, Comp. II, 
Comp. III  respectively, given in Tab.~\ref{tab:elem}.}
\label{fig3a:iron}
\end{figure}

Figures~~\ref{fig2:spec}~-~\ref{fig6:fe} present outgoing intensity 
spectra of the whole accretion disk at various aspect angles. The spectra
were obtained by integration over eight rings located at different radii,
see Tab.~\ref{tab:disk}. Particular lines in 
Figs.~\ref{fig2:spec}~-~\ref{fig6:fe}
correspond to different angles between the direction to the observer and
the normal to the disk. Exact values of those eight angles and their cosines
are given in Table~\ref{tab:cosin}. In further discussion we draw 
attention of the reader to the extreme angles: solid black line represents
almost vertical direction (face-on aspect), whereas the solid red
line represents almost horizontal direction (edge-on aspect).  

The spectrum of our canonical model is compared with the model 
of the same spectral index but different X-ray luminosities,
Figs.~\ref{fig2:spec}, and~\ref{fig4:lum}, and of the same X-ray luminosity
but different spectral indices Fig.~\ref{fig3:bb}, and~\ref{fig5:ind}.
Spectra for various chemical compositions and the same irradiation
are plotted in Figs.~\ref{fig3a:iron}, and~\ref{fig6:fe}.

Panel (a) in Fig.~\ref{fig2:spec} presents intensity spectra of the disk 
with no irradiation. Comparison to panels (b)-(d) shows, that all emerging
radiation  above above 0.2 keV originates from the disk
atmosphere as the result of an external hard X-ray illumination. 

In all irradiated models outgoing spectra show the effect, that the intensity
of radiation emerging almost horizontally to the disk
($i=88.9^{\circ}$) is greater than the intensity emerging almost
vertically ($i=11.4^{\circ}$). This is valid in the X-ray energy
band from about 0.1 keV up to 100 keV, and in the infrared band. 
Such an effect of 
limb-brightening should be attributed to the inversion of temperature in
the irradiated disk atmospheres, see Fig.~\ref{fig1:struc}.

All broad-band spectra show the deficit of hard X-rays, which were
absorbed in the disk and suffered continuum Compton downscattering. 
This caused that the energy of hard X-rays  was redistributed to lower 
energies. Moreover, there exists a spectral bump below 1 keV, which 
is similar to the well known soft X-ray excess. Such bump is
more prominent for the disk seen edge on (red solid line) in all
models. 

\section{Iron line complex}
\label{res:line}

External irradiation generates two iron fluorescent lines:  $K_\alpha$
doublet at 6.391 and 6.404 keV, and $K_\beta$ singlet  at 7.057 keV. 
Furthermore, 
due to Compton scattering of the line photons 
in the irradiated atmosphere photons from line core 
are shifted toward lower energies creating red wing, which is 
called Compton shoulder. All the above features are simultaneously 
reproduced in our model calculations.

\begin{figure*}
\epsfxsize=8.8cm \epsfbox[175 450 445 700]{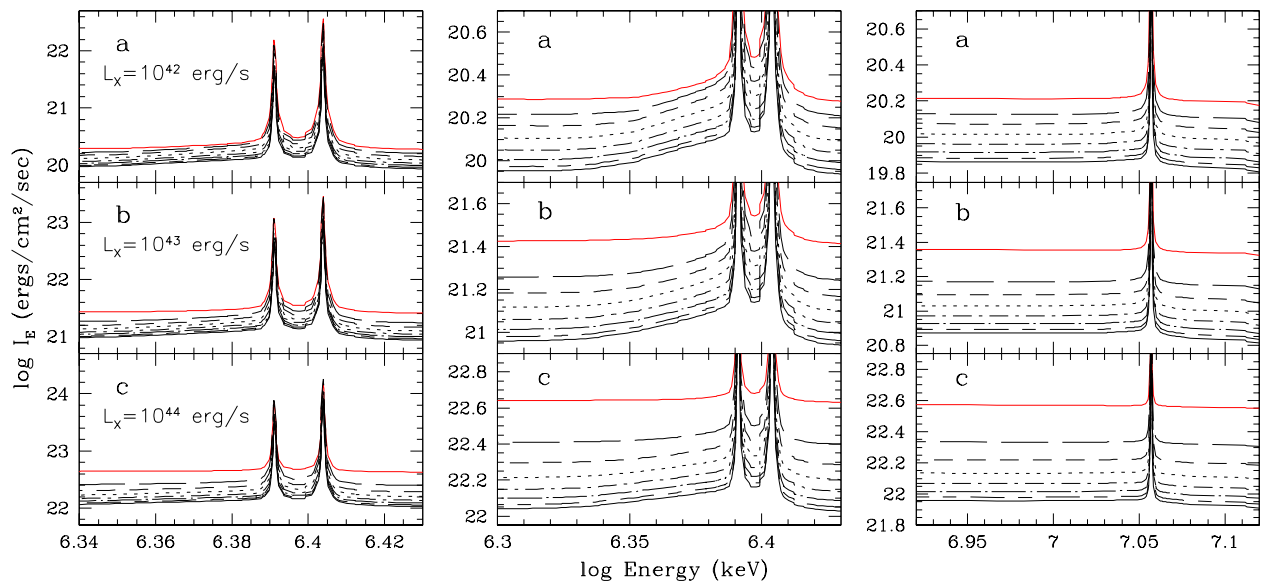}
\caption{Iron line complex in the disk spectra integrated over radii for 8
different aspect angles (see Table~\ref{tab:cosin}). 
Panels (a), (b),and  (c) show model spectra
with the same $\alpha_X=0.9$, but for three different
luminosities $10^{42}$, $10^{43}$ and $10^{44}$ erg~s$^{-1}$, respectively. 
Iron $K_\alpha$ doublet is presented in the left panels, 
whereas Compton shoulder of this line is shown in the middle panels. 
Right panels show $K_\beta$ line with Compton shoulders of a
marginal height. }
\label{fig4:lum}
\end{figure*}

\begin{figure*}
\epsfxsize=8.8cm \epsfbox[175 450 445 700]{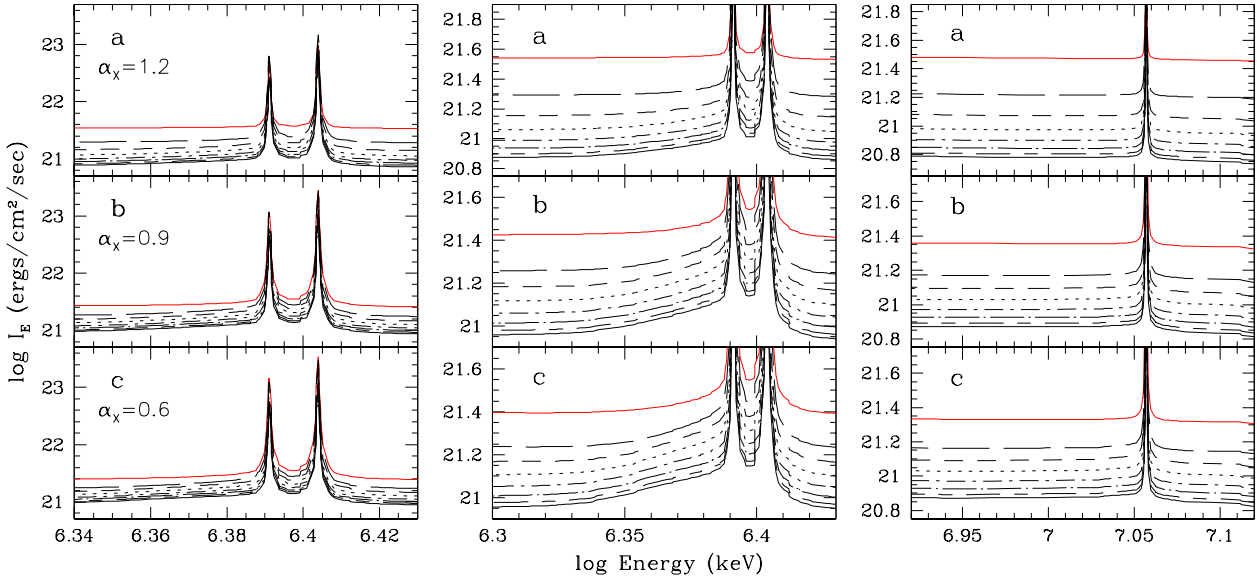}
\caption{Iron line complex in the disk spectra integrated over radii for 8
different aspect angles (see Table~\ref{tab:cosin}).  
Panels (a), (b),and  (c) show model spectra 
with the same $L_X=10^{43}$ erg~s$^{-1}$, but for three different
spectral indices $1.2$, $0.9$ and $0.6$, respectively. 
Iron $K_\alpha$ doublet is presented in the left panels,
whereas Compton shoulder of this line is shown in the middle panels.
Right panels show  $K_\beta$ line with very low red-wing Compton shoulders.}
\label{fig5:ind}
\end{figure*}

\begin{figure*}
\epsfxsize=8.8cm \epsfbox[175 450 445 700]{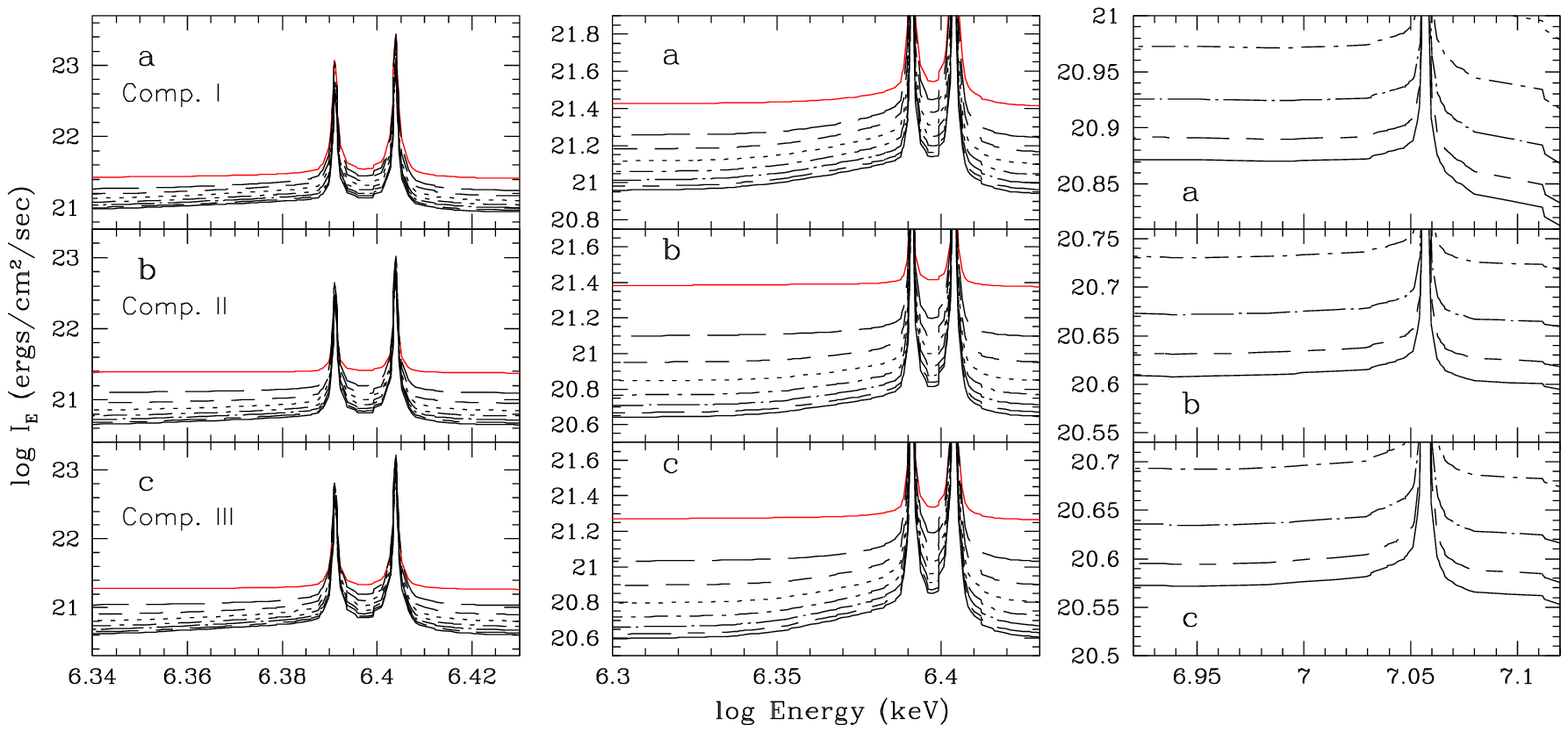}
\caption{Iron line complex in the disk spectra integrated over radii for 8
different aspect angles (see Table~\ref{tab:cosin}).  
Panels (a), (b),and  (c) show model spectra 
with the same $L_X=10^{43}$ erg~s$^{-1}$, and spectral
index $\alpha_X=0.9$, but for three different
chemical compositions, Comp. I, Comp. II, and Comp. III respectively. 
Iron $K_\alpha$ doublet is presented in the left panels,
whereas Compton shoulder of this line is shown in the middle panels.
Right panels show enlarged $K_\beta$ lines with their weak Compton shoulders.}
\label{fig6:fe}
\end{figure*}

Figs.~\ref{fig4:lum},~\ref{fig5:ind},~\ref{fig6:fe} present 
integrated intensity spectra just around the fluorescent $K_\alpha$ doublet
(left panels), Compton shoulder (middle panels), and fluorescent
$K_\beta$ singlet (right panels) for all computed spectra. 
Unfortunately, Compton shoulder of $K_\beta$ line is very weak in
our models. 

Limb-brightening is clearly seen in the region of iron line complex. 
The Compton shoulder is most prominent in 
the vertical direction (black solid line), while it almost disappears 
for the disk seen edge on (red solid line). 

\begin{table*}
\begin{center}
\caption{Equivalent widths of modeled iron line components: $K_\alpha$ ,
its Compton Shoulder ($CS_\alpha$), and $K_\beta$, and its Compton
Shoulder ($CS_\beta$). All values are given in eV for two extreme
aspect angles: $i=11.4^{\circ}$ - face-on, and $i=88.9^{\circ}$ - edge-on. }
\label{tab:eqw}
\begin{tabular}{cclccccccccc} 
\hline \hline 
  &    &   Chem.&$K_\alpha$ & $K_\alpha$ & $CS_\alpha$ & $CS_\alpha$ &
  $K_\beta$ & $K_\beta$ & $CS_\beta$ & $CS_\beta$ \\
$\alpha_X$ & $\log(L_X) $ & copm. & $i=11.4^{\circ}$& $i=88.9^{\circ}$ 
& $i=11.4^{\circ}$ &  $i=88.9^{\circ}$ & $i=11.4^{\circ}$ & $i=88.9^{\circ}$ 
        & $i=11.4^{\circ}$ & $i=88.9^{\circ}$ \\ 
0.9 &  42 & Comp. I  & 104.31  & 211.53 & 15.34  & 6.34 & 15.61 & 31.68 & 1.39 & 0.51\\
0.9 &  43 & Comp. I  & 108.18  & 115.33 & 12.13 & 4.36  & 15.39 & 16.90 & 1.24& 0.27\\
0.9 &  44 & Comp. I &  68.11 &  39.44  &  9.22 & 2.12 & 9.58 & 5.83 & 0.97&0.18 \\
0.6 &  43 & Comp. I & 111.12 & 144.25 & 13.33 & 3.73  & 15.75 & 20.85 & 1.36 & 0.40\\
1.2 &  43 & Comp. I &  82.66 & 30.65 & 6.92 & 1.55  & 11.96 & 4.65 & 0.76 & 0.14\\
0.9 &  43 & Comp. II & 125.06 &30.10 &6.57 & 1.78 &16.67 & 4.88 & 0.57& 0.06\\
0.9 &  43 & Comp. III & 210.15 & 67.04& 9.88 & 3.14& 27.92& 10.89& 0.72& 0.00\\
\hline \hline
\end{tabular}
\end{center}
\end{table*}

The amount of energy emitted by iron line features depends on the shape 
and luminosity of irradiating continuum and on its spectral index.  
We have computed equivalent widths of modeled iron line components,
and collected them in Table~\ref{tab:eqw}  for two extreme aspect angles:
face on ($i=11.4^{\circ}$) and edge on, $i=88.9^{\circ}$.
Features are stronger for models with lower luminosity and 
harder spectral index.  For such cases, also  $K_\alpha$
and $K_\beta$ lines seen edge on are 
more prominent. The energy emitted in Compton shoulder is about ten and
more times lower than the energy emitted in $K_\alpha$ line.

\subsection{Chemical composition of the accretion disk}

Presence of iron atoms in the gas of Comp. I ensures generation of
iron $K_\alpha$ and $K_\beta$ iron fluorescent lines and their Compton 
Shoulders (Tab.~\ref{tab:eqw} first five rows). 
Both fluorescent lines originate in deeper (colder) atmospheric
layers, where iron atoms are only weakly ionized. However, if 
the actual  chemical composition is rich of various  heavy elements,
like C, O, Si, S etc., then the appearance and equivalent widths
of both fluorescent lines can be reduced due to K-edge absorption of 
X-ray photons by all those elements. 

To check this we have computed disk models for our canonical 
X-ray luminosity $L_X=10^{43}$ erg s$^{-1}$ and spectral index
$\alpha_{X}=0.9$ assuming other chemical compositions (Comp. II and
Comp. III) with the excess of helium and the solar chemical abundance
of eight heavier elements: C, N, O, Ne, Mg, Si, S and Fe 
(see Table.~\ref{tab:elem}). 

Table.~\ref{tab:eqw} shows that the equivalent widths of the Compton
Shoulder has the highest value in models of Comp. I, both 
for $K_\alpha$ and $K_\beta$ lines. Addition of heavy elements 
other than iron (Comp. II, and Comp. III) and their K-edge absorption, 
significantly decreases EW of Compton Shoulder 
(compare rows 2, 6 and 7 of Table.~\ref{tab:eqw}).  
 The same holds for $K_{\alpha}$ and $K_{\beta}$ line cores seen
edge-on. But for the disk seen face-on ($i=11.4^{\circ}$), the situation 
for line cores is opposite. EWs of both line cores increase with 
addition of heavy elements.

\section{Discussion}
\label{sec:discussion}

Calculations of an accretion disk atmosphere irradiated by external hard
X-rays are very time consuming and sometimes converge with difficulty. 
In our paper we have computed the disk up
to the distance from the central black hole equal $18 R_{Schw}$. 
Model atmospheres of farther rings did not 
converge with our computer code. 
Going farther from the black hole, the gravity in the disk atmosphere
becomes lower, and irradiation disturbs hydrostatic
equilibrium due to strong radiative pressure. 
This problem should be solved using dynamical calculations, but 
none of recently available hydrodynamic codes contains such 
complicated radiative transfer as presented in this paper. 

Our treatment of Compton scattering allowed us to compute 
the structure of the Compton
heated skin in an accretion disk atmosphere together with the transition
layer down to the deep cold zone \citep{hubeny2001}. 
Structure of the  transition layer depends on heavy
elements abundance. We have shown that 
the position of a small horizontal branch around $3 \times 10^5$~K 
changes with chemical composition of an irradiated gas.
 
In the soft X-ray domain of our model spectra we observe huge 
increase of specific intensity due to Compton redistribution of hard X-ray
photons toward lower energies. We claim here, that those bumps 
give impact to soft X-ray excess frequently observed in many objects. 
Such excess in our models is higher for higher luminosity of
X-ray source and for softer (i.e. higher) power-law index. 

All spectra discussed here were computed in the frame of the accretion
disk, so we  did not consider line broadening and distortion caused by
Keplerian motion of gas around black hole.  
Our goal was to demonstrate the existence of the red wing due to Compton 
scattering of fluorescent line photons on a colder electron gas. 
Narrow $K_\alpha$ lines recently observed by {\it XMM} 
{\it Chandra} and {\it Suzaku} satellites have equivalent width 
75 eV in case of MCG+8-11-11 \citep{matt2006}, 165 eV in NGC 2992
\citep{yaqoob2007} and 190 eV in the famous object MCG-6-30-15 
\citep{miniutti2007}.
Those values of EWs are reproduced by our model calculations,
 see Table~\ref{tab:eqw}.
The resolution of actually working satellites is unable to resolve iron
$K_\alpha$ double profile. 

Similar comparison with $K_\beta$ line shows excellent agreement of our 
models with observations. \citet{yaqoob2007} reported detection of 
$K_\beta$ iron line in NGC 2992 with an EW of 
about 27 eV, and \citet{matt2006}
estimated an upper limit of 10 eV in  MCG+8-11-11. 
 
There are not many detections of Compton shoulder in AGN. Most probably,
because it is very difficult to distinguish between red wing produced
by Compton scattering and that caused by  relativistic disk motions. 
Nevertheless, EW of Compton shoulder fitted by \citet{matt2006} in 
MCG+8-11-11 equals $17\pm 10$ eV.  Such a value 
agrees with values obtain in our computations for the disk seen face
on, which contains only hydrogen, helium and iron (Comp. I). 
EW and the profile of Compton shoulder presented in this paper 
were computed on the bases of detailed model atmosphere calculations,
whereas  \citet{matt2002} estimated them from models of irradiated 
``wall'' of constant density. 

Our models of broad-band intensity spectra  from the irradiated
accretion disk are interesting also in other spectral domains,
for example in infrared and optical bands.
They can be immediately used to study other important spectral features
like hydrogen Lyman edge which looks different depending on aspect
angle. We address this issue for future consideration.    

\section{Summary}
\label{sec:sum}

In this paper we presented intensity spectra emerging from a sample 
accretion disk around a supermassive black hole in AGN. The disk was 
irradiated by hard X-rays of the power-law spectral shape 
originating from a point-like source located above the inner disk.
We compared results for the cases when irradiation luminosity,
power-law spectral index, and chemical composition change. 

Reprocessing of external hard X-rays causes generation of a very hot
external layer in the disk atmosphere where temperature rises to millions K,
whereas the structure of deeper atmospheric layers is not affected at all. 

We numerically reproduced redistribution of external X-rays both
by thermal emission from the heated disk skin and by Compton downscattering
 in deeper and colder layers of the disk atmosphere. As the result
outgoing intensity spectra of the disk exhibit the following important 
observable properties.

First, external irradiation generates two fluorescent lines of iron, 
$K_\alpha$ and $K_\beta$. Compton scattering of the line radiation on 
colder electron gas (below the hot disk skin) generates the 
 Compton shoulder on the red side of both $K_\alpha$ and  
$K_\beta$ fluorescent
lines, nevertheless shoulder is very weak for $K_\beta$ line.
Compton shoulders modeled here differs from the Compton shoulder presented
by \citet{matt2002}, when the latter resulted from X-ray reflection off the
constant density slab. Equivalent widths of iron line complex agree with 
recent observations \citep{matt2006,yaqoob2007,miniutti2007}.

Second, Compton scattering of the external X-rays causes build up of
the soft X-ray excess below 1 keV. This excess is higher for the disk 
observed face on, but still too low to explain observations. 

Third, our results clearly demonstrate the effect of limb brightening
of the disk radiation mostly in the X-ray domain. This result can 
put new constrains on the geometry of the X-ray source and the reflecting 
disk atmosphere. We address this issue for further consideration. 

\section*{Acknowledgments}

This research was supported by the Polish Committee for Scientific
Research grant No. N N203 4061 33

\bibliographystyle{mn2e}
\bibliography{myrefs2}

\begin{thebibliography}{}

\bibitem[\protect\citeauthoryear{{Ballantyne}, {Ross} \& {Fabian}}{{Ballantyne}
  et~al.}{2001}]{ballantyne2001}
{Ballantyne} D.~R.,  {Ross} R.~R.,    {Fabian} A.~C.,  2001, MNRAS, 327, 10

\bibitem[\protect\citeauthoryear{{Basko}}{{Basko}}{1978}]{basko1978}
{Basko} M.~M.,  1978, ApJ, 223, 268

\bibitem[\protect\citeauthoryear{{Chaisson} \& {McMillan}}{{Chaisson} \&
  {McMillan}}{1996}]{chaisson96}
{Chaisson} E.,  {McMillan} S.,  1996, {Astronomy today}.
Upper Saddle River, N.J.~: Prentice Hall, c1996.~2nd ed.

\bibitem[\protect\citeauthoryear{{Done}, {Madejski}, {{\.Z}ycki} \&
  {Greenhill}}{{Done} et~al.}{2003}]{done2003}
{Done} C.,  {Madejski} G.~M.,  {{\.Z}ycki} P.~T.,    {Greenhill} L.~J.,  2003,
  ApJ, 588, 763

\bibitem[\protect\citeauthoryear{{Fabian}, {Rees}, {Stella} \&
  {White}}{{Fabian} et~al.}{1989}]{fabian1989}
{Fabian} A.~C.,  {Rees} M.~J.,  {Stella} L.,    {White} N.~E.,  1989, MNRAS,
  238, 729

\bibitem[\protect\citeauthoryear{{Guainazzi}, {Perola}, {Matt}, {Nicastro},
  {Bassani}, {Fiore}, {dal Fiume} \& {Piro}}{{Guainazzi}
  et~al.}{1999}]{guainazzi1999}
{Guainazzi} M.,  {Perola} G.~C.,  {Matt} G.,  {Nicastro} F.,  {Bassani} L.,
  {Fiore} F.,  {dal Fiume} D.,    {Piro} L.,  1999, A\&A, 346, 407

\bibitem[\protect\citeauthoryear{{Guilbert}}{{Guilbert}}{1981}]{guilbert1981}
{Guilbert} P.~W.,  1981, MNRAS, 197, 451

\bibitem[\protect\citeauthoryear{{Hubeny}, {Blaes}, {Krolik} \&
  {Agol}}{{Hubeny} et~al.}{2001}]{hubeny2001}
{Hubeny} I.,  {Blaes} O.,  {Krolik} J.~H.,    {Agol} E.,  2001, ApJ, 559, 680

\bibitem[\protect\citeauthoryear{{Iwasawa}, {Fabian}, {Young}, {Inoue} \&
  {Matsumoto}}{{Iwasawa} et~al.}{1999}]{iwasawa1999}
{Iwasawa} K.,  {Fabian} A.~C.,  {Young} A.~J.,  {Inoue} H.,    {Matsumoto} C.,
  1999, MNRAS, 306, L19

\bibitem[\protect\citeauthoryear{{Krause} \& {Oliver}}{{Krause} \&
  {Oliver}}{1979}]{krause1979}
{Krause} M.~O.,  {Oliver} J.~H.,  1979, Journal of Physical and Chemical
  Reference Data, 8, 329

\bibitem[\protect\citeauthoryear{{Madej}}{{Madej}}{1991}]{madej1991}
{Madej} J.,  1991, ApJ, 376, 161

\bibitem[\protect\citeauthoryear{{Madej} \& {R{\'o}{\.z}a{\'n}ska}}{{Madej} \&
  {R{\'o}{\.z}a{\'n}ska}}{2000a}]{madej2000a}
{Madej} J.,  {R{\'o}{\.z}a{\'n}ska} A.,  2000a, A\&A, 356, 654

\bibitem[\protect\citeauthoryear{{Madej} \& {R{\'o}{\.z}a{\'n}ska}}{{Madej} \&
  {R{\'o}{\.z}a{\'n}ska}}{2000b}]{madej2000b}
{Madej} J.,  {R{\'o}{\.z}a{\'n}ska} A.,  2000b, A\&A, 363, 1055

\bibitem[\protect\citeauthoryear{{Madej} \& {R{\'o}{\.z}a{\'n}ska}}{{Madej} \&
  {R{\'o}{\.z}a{\'n}ska}}{2004}]{madej2004}
{Madej} J.,  {R{\'o}{\.z}a{\'n}ska} A.,  2004, MNRAS, 347, 1266

\bibitem[\protect\citeauthoryear{{Markowitz}, {Takahashi}, {Watanabe},
  {Nakazawa}, {Fukazawa}, {Kokubun}, {Makishima}, {Awaki}, {Bamba} \&
  {Isobe}}{{Markowitz} et~al.}{2007}]{markowitz2007}
{Markowitz} A.,  {Takahashi} T.,  {Watanabe} S.,  {Nakazawa} K.,  {Fukazawa}
  Y.,  {Kokubun} M.,  {Makishima} K.,  {Awaki} H.,  {Bamba} A.,    {Isobe} N.,
  2007, ApJ, 665, 209

\bibitem[\protect\citeauthoryear{{Matt}}{{Matt}}{2002}]{matt2002}
{Matt} G.,  2002, MNRAS, 337, 147

\bibitem[\protect\citeauthoryear{{Matt}, {Bianchi}, {de Rosa}, {Grandi} \&
  {Perola}}{{Matt} et~al.}{2006}]{matt2006}
{Matt} G.,  {Bianchi} S.,  {de Rosa} A.,  {Grandi} P.,    {Perola} G.~C.,
  2006, A\&A, 445, 451

\bibitem[\protect\citeauthoryear{{Mihalas}}{{Mihalas}}{1978}]{mihalas1978}
{Mihalas} D.,  1978, {Stellar atmospheres /2nd edition/}.
San Francisco, W.~H.~Freeman and Co., p.~650

\bibitem[\protect\citeauthoryear{{Miniutti}, {Fabian}, {Anabuki}, {Crummy},
  {Fukazawa}, {Gallo}, {Haba}, {Hayashida} \& {Holt}}{{Miniutti}
  et~al.}{2007}]{miniutti2007}
{Miniutti} G.,  {Fabian} A.~C.,  {Anabuki} N.,  {Crummy} J.,  {Fukazawa} Y.,
  {Gallo} L.,  {Haba} Y.,  {Hayashida} K.,    {Holt} S.,  2007, PASJ, 59, 315

\bibitem[\protect\citeauthoryear{{Nandra} \& {Iwasawa}}{{Nandra} \&
  {Iwasawa}}{2007}]{nandra2007}
{Nandra} K.,  {Iwasawa} K.,  2007, MNRAS, 382, L1

\bibitem[\protect\citeauthoryear{{Nayakshin} \& {Kallman}}{{Nayakshin} \&
  {Kallman}}{2001}]{nayakshin2001}
{Nayakshin} S.,  {Kallman} T.~R.,  2001, Apj, 546, 406

\bibitem[\protect\citeauthoryear{{Nayakshin}, {Kazanas} \&
  {Kallman}}{{Nayakshin} et~al.}{2000}]{nayakshin2000}
{Nayakshin} S.,  {Kazanas} D.,    {Kallman} T.~R.,  2000, ApJ, 537, 833

\bibitem[\protect\citeauthoryear{{Petrucci}, {Ponti}, {Matt}, {Longinotti},
  {Malzac}, {Mouchet}, {Boisson}, {Maraschi}, {Nandra} \&
  {Ferrando}}{{Petrucci} et~al.}{2007}]{petrucci2007}
{Petrucci} P.~O.,  {Ponti} G.,  {Matt} G.,  {Longinotti} A.~L.,  {Malzac} J.,
  {Mouchet} M.,  {Boisson} C.,  {Maraschi} L.,  {Nandra} K.,    {Ferrando} P.,
  2007, A\&A, 470, 889

\bibitem[\protect\citeauthoryear{{Pounds}, {Nandra}, {Stewart}, {George} \&
  {Fabian}}{{Pounds} et~al.}{1990}]{pounds1990}
{Pounds} K.~A.,  {Nandra} K.,  {Stewart} G.~C.,  {George} I.~M.,    {Fabian}
  A.~C.,  1990, Nature, 344, 132

\bibitem[\protect\citeauthoryear{{Reeves}, {Awaki}, {Dewangan}, {Fabian},
  {Fukazawa}, {Gallo}, {Griffiths}, {Inoue} \& {Kunieda}}{{Reeves}
  et~al.}{2007}]{reeves2007}
{Reeves} J.~N.,  {Awaki} H.,  {Dewangan} G.~C.,  {Fabian} A.~C.,  {Fukazawa}
  Y.,  {Gallo} L.,  {Griffiths} R.,  {Inoue} H.,    {Kunieda} H.,  2007, PASJ,
  59, 301

\bibitem[\protect\citeauthoryear{{Reeves}, {Nandra}, {George}, {Pounds},
  {Turner} \& {Yaqoob}}{{Reeves} et~al.}{2004}]{reeves2004}
{Reeves} J.~N.,  {Nandra} K.,  {George} I.~M.,  {Pounds} K.~A.,  {Turner}
  T.~J.,    {Yaqoob} T.,  2004, ApJ, 602, 648

\bibitem[\protect\citeauthoryear{{Reynolds} \& {Begelman}}{{Reynolds} \&
  {Begelman}}{1997}]{reynolds1997}
{Reynolds} C.~S.,  {Begelman} M.~C.,  1997, ApJ, 488, 109

\bibitem[\protect\citeauthoryear{{Risaliti}, {Bianchi}, {Matt}, {Baldi},
  {Elvis}, {Fabbiano} \& {Zezas}}{{Risaliti} et~al.}{2005}]{risaliti2005}
{Risaliti} G.,  {Bianchi} S.,  {Matt} G.,  {Baldi} A.,  {Elvis} M.,  {Fabbiano}
  G.,    {Zezas} A.,  2005, ApJ, 630, L129

\bibitem[\protect\citeauthoryear{{R{\'o}{\.z}a{\'n}ska}, {Czerny}, {{\.Z}ycki}
  \& {Pojma{\'n}ski}}{{R{\'o}{\.z}a{\'n}ska} et~al.}{1999}]{rozanska1999}
{R{\'o}{\.z}a{\'n}ska} A.,  {Czerny} B.,  {{\.Z}ycki} P.~T.,    {Pojma{\'n}ski}
  G.,  1999, MNRAS, 305, 481

\bibitem[\protect\citeauthoryear{{R{\'o}{\.z}a{\'n}ska}, {Dumont}, {Czerny} \&
  {Collin}}{{R{\'o}{\.z}a{\'n}ska} et~al.}{2002}]{rozanska2002}
{R{\'o}{\.z}a{\'n}ska} A.,  {Dumont} A.-M.,  {Czerny} B.,    {Collin} S.,
  2002, MNRAS, 332, 799

\bibitem[\protect\citeauthoryear{{Tanaka}, {Nandra}, {Fabian}, {Inoue},
  {Otani}, {Dotani}, {Hayashida}, {Iwasawa}, {Kii}, {Kunieda}, {Makino} \&
  {Matsuoka}}{{Tanaka} et~al.}{1995}]{tanaka1995}
{Tanaka} Y.,  {Nandra} K.,  {Fabian} A.~C.,  {Inoue} H.,  {Otani} C.,  {Dotani}
  T.,  {Hayashida} K.,  {Iwasawa} K.,  {Kii} T.,  {Kunieda} H.,  {Makino} F.,
   {Matsuoka} M.,  1995, Nature, 375, 659

\bibitem[\protect\citeauthoryear{{Yaqoob}, {Murphy}, {Griffiths}, {Haba},
  {Inoue}, {Itoh}, {Kelley}, {Kokubun}, {Markowitz}, {Mushotzky}, {Okajima},
  {Ptak}, {Reeves}, {Serlemitsos}, {Takahashi} \& {Terashima}}{{Yaqoob}
  et~al.}{2007}]{yaqoob2007}
{Yaqoob} T.,  {Murphy} K.~D.,  {Griffiths} R.~E.,  {Haba} Y.,  {Inoue} H.,
  {Itoh} T.,  {Kelley} R.,  {Kokubun} M.,  {Markowitz} A.,  {Mushotzky} R.,
  {Okajima} T.,  {Ptak} A.,  {Reeves} J.,  {Serlemitsos} P.~J.,  {Takahashi}
  T.,    {Terashima} Y.,  2007, PASJ, 59, 283

\bibitem[\protect\citeauthoryear{{Yaqoob} \& {Padmanabhan}}{{Yaqoob} \&
  {Padmanabhan}}{2004}]{yaqoob2004}
{Yaqoob} T.,  {Padmanabhan} U.,  2004, ApJ, 604, 63

\end{thebibliography}

\end{document}